\documentclass[vecphys]{svmult}

\usepackage{makeidx}         
\usepackage{graphicx}        
\usepackage{multicol}        
\usepackage[bottom]{footmisc}

\makeindex             

\begin{document}

\title*{Very Large Spectroscopic Surveys with the VLT}
\author{I.\ R.\ Parry\inst{1}}
\authorrunning{I.\ R.\ Parry}
\institute{Institute of Astronomy, Madingley Road, Cambridge, CB3 0HA, United Kingdom \texttt{irp@ast.cam.ac.uk}}
%
%
\maketitle

Recently, it has been recognised that very large spectroscopic surveys
(several million spectra) are required to advance our understanding of 
Dark Energy (via baryonic wiggles) and the detailed history of our Local Group
of galaxies (via Galactic Archaeology or near-field cosmology). In this paper
I make a preliminary exploration of how this might be done by putting a wide
field, optical, prime-focus fibre-fed spectroscopic facility on one of the
VLT's UTs.

\section{Introduction}
\label{sec:1}

The Gemini Aspen process identified the need for a wide field (1-2 degree diameter),
optical, multi-object (few thousand at a time) spectroscopic survey facility
to study Dark Energy (DE) and Galactic Archaeology (GA).
A feasibility study was completed in Mar 2005 and following that, two competing
design studies are currently being carried out for a system which will go at
prime focus on the Japanese Subaru Telescope. This proposed system is called WFMOS (Wide Field
Multi-Object Spectrograph).

Four potential surveys were identified. 1) A low redshift DE survey, 2) a high
redshift DE survey, 3) A low-resolution GA survey and 4) a high resolution
GA survey. These require a total of about 7 million spectra and about 600
nights of observing.

Although the Gemini design study process is well underway it is still worth considering doing a similar
project on the VLT for several important reasons. The GA case is better served from the southern
hemisphere which gives access to the galactic bulge and the Magellanic Clouds. It is
very likely that most future large investments in ground based facilities (i.e. ELTs)
will be in the South and this instrument will be an important path-finder for ELT follow up studies.
At the time of writing, no European astronomers have access to Gemini.
The facility could be used for science other than DE and GA (basically it's about 50
times faster than the current FLAMES instrument). On the VLT one of the UTs could be 80\% - 90\% dedicated
to this project (the remainder going to VLTI) and the surveys could be done in about
2 years - this cannot happen at Subaru where the
fraction of time given to WFMOS would be much less.
WFMOS is a design study and it's not 100\% certain that it will be fully funded and continue
in to the construction phase.

In this paper I present a very preliminary discussion of the technical,
operational and cost implications of putting something like WFMOS on the VLT.

\section{Instrumental Requirements}

The science case for the dark energy and galactic archaeology surveys
demands $\sim$2400 (or more) simultaneous spectra at optical wavelengths
for discrete objects in a 1.5 deg FOV on an 8m telescope with perhaps
up to 20 field configurations per night.
The large FOV, the large wavelength coverage and the high spectral
resolution are best addressed by a fibre-fed multi-object spectroscopy (MOS) approach with one
fibre per object.

The field change time therefore has to be fast, $<5$ minutes from the end
of one exposure to the start of another one on a new field.
Up to $2,400\times20=48,000$ fibres have to be placed in a 24 hour cycle
(for one of the DE surveys). Each fibre tip ideally has to be placed accurately
with 5 degrees of freedom: X, Y	for the object position, Z for focus and tip
and tilt for pupil aiming.

The Galactic Archaeology high resolution survey drives the spectrograph design.
At a spectral resolution R$\sim$20,000, in the wavelength range 480nm to 680nm, a spectral resolution element
(SRE) is 0.029nm.
One spectrum has 6897 SREs, which is 245,760 detector pixels for a single spectrum
(assuming an f/1.8 camera and allowing for gaps between the spectra).
Given the surface density of targets 800 simultaneous spectra are needed which requires
at least 197 million detector pixels. A 16k$\times$12k detector format has 201 million pixels.

These numbers also assume that the f-ratio into the fibres is about f/2.3, from the prime
focus corrector (PFC) and the f-ratio of the collimator is about f/2.0 to allow for
focal ratio degradation. A fibre diameter of about 92 microns (1 arcsec) is assumed and the 
201 Mega-pixel detector format is acheived with 24 chips each with 2048$\times$4096,
15 micron pixels.
A camera f-ratio of f/1.8 is assumed. 
A large beam diameter is required to achieve the high spectral resolution of R$\sim$20,000.
The spectrograph(s) will therefore be very large and will have to be located on
one or both Nasmyth platforms.

The fibre positioning system is a major challenge. Magnetically held fibres
(OzPoz/2df/Autofib type systems) cannot really cope with the large numbers required. 
Fishermen round the pond systems (like MX and KMOS) also cannot cope with the large numbers.
Systems which have a micro-positioner per fibre (e.g. like the Echidna system on
FMOS/Subaru) are being investigated as part of the WFMOS design study and can feasibly
deliver the $\sim$2400 fibres required in the FOV which is about 500mm in diameter.
Plug-plate systems have the greatest potential in terms of the number of simultaneous on-sky fibres
because they have the minimal amount of hardware associated with each fibre tip.

\begin{figure}[t]
\centering
\includegraphics[height=10cm]{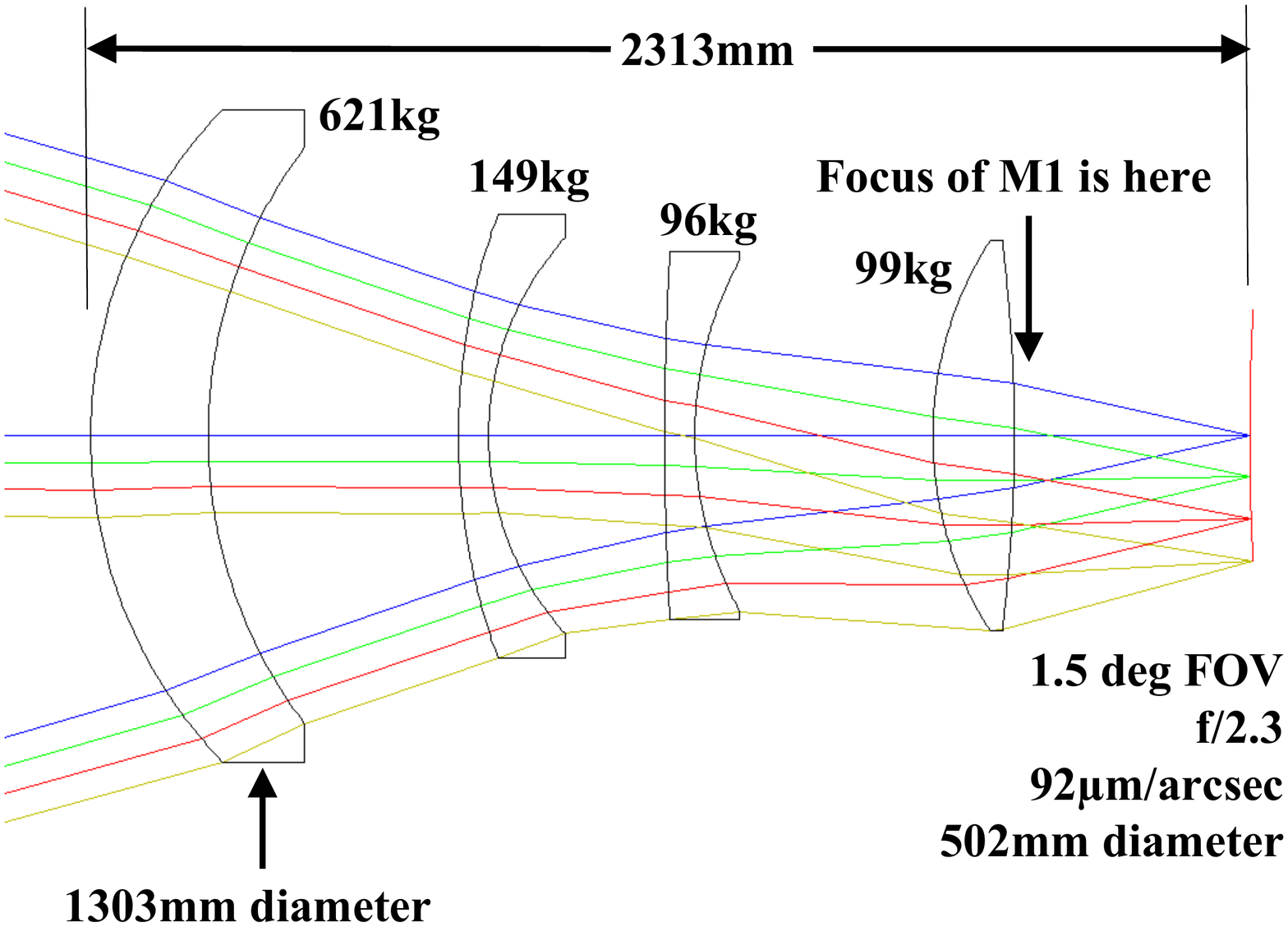}
%
%
\caption{Preliminary design of a PFC for the VLT. The 1.5 degree
FOV is unvignetted (unlike the WFMOS/Subaru design). The design lacks
an atmospheric dispersion compensator.}
\label{parryfig:1}       
\end{figure}

A very preliminary PFC design is shown in figure \ref{parryfig:1}. This design delivers excellent image quality
(point source spots smaller than 0.25 arcsec) over the whole FOV. However, the mass of the glass alone in this
design is 965kg.
Furthermore, it needs an atmospheric dispersion corrector. More work is required to optimise
the design in terms of it's mass, telecentric properties, and number of aspheric surfaces.
It may well be possible to increase the FOV above 1.5 degrees but this will have significant
mass and cost implications. Although this design is not optimal it does give a good estimate of
the size and location of the focal plane and the location of the PFC optics.

\begin{figure}[t]
\centering
\includegraphics[width=10cm]{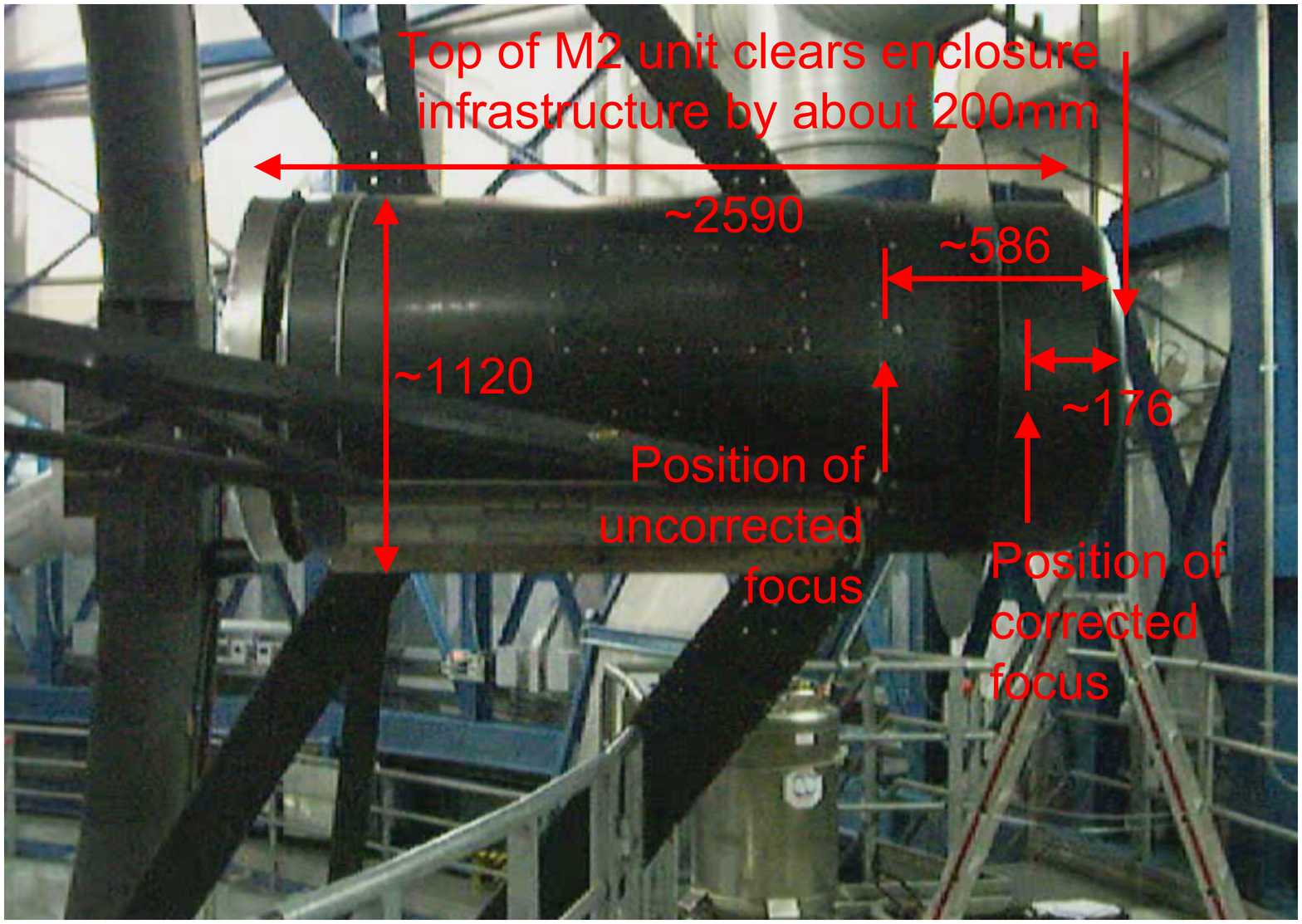}
\caption{The M2 unit on UT4 showing some key dimensions and masses.}
\label{parryfig:2}
\end{figure}

Figure \ref{parryfig:2} shows the existing M2 unit (secondary mirror unit) on UT4 with approximate dimensions.
This shows that there is very little space for a fibre positioning system given the potential
for collisions with the existing enclosure infrastructure. If the mass of the existing M2 unit
($\sim1700$ kg) cannot be exceded then it will also be hard to fit a PFC, an aquisition and
guidance system (including the optics to control the figure of the primary mirror) and a fibre
positioning system at prime and stay within the mass budget. 

\section{A possible plug-plate solution to the fibre positioning problem}

In this scheme the spectrographs and most of the length of the optical fibre bundle are
permanently mounted on the telescope. The long fibre bundle is terminated at the top with an efficient, multi-way
fibre-optic connector. The plugplate itself is basically a sheet of suitable material with accurately drilled holes into which the fibres are inserted.
The plugplate modules are configured off the telescope.
With exposures of $\sim$30 minutes, up to 20 pre-configured modules have to be ready for use at the start of a night (unlikely to be more than this).
A special machine interchanges the modules on the telescope during the night: for each observation a pre-configured fibre module is mounted at the focus
and connected to the fibre bundle.
In each module the fibres are short ($\sim$500-1000mm) and the output ends are arranged in a fixed pattern (a fibre-optic connector).

The operation of the system is as follows.
At the end of an exposure the telescope moves in altitude and azimuth so that the prime
focus is at the access platform where a specialised machine (the module interchanger)
can access the fibre module.
The permanently installed part of the fibre connector is backed off from the connector
half on the plugplate module.
The used module is then removed and the next one is put on. Next, the new module is
referenced to the telescope focal plane
The permanently installed part of the fibre connector then mates with the other half
of the connector on the module.
The telescope slews to the next field.
This entire process is the same for all modules.

The fibre-modules are prepared in a dedicated space, probably at the base camp.
The plates are drilled on CNC machines. The facility has enough machines to accurately
drill up to 48,000 holes in an 8 hour shift. For example, 20 machines are needed if the
time to drill one hole is 12 seconds.
The plates are non-metallic to reduce costs and reduce fitting tolerances.
The fibres are removed from used plates and inserted into new plates by either robots or people.
The facility has enough fibre-pluggers to reconfigure 48,000 fibres in an 8 hour shift.
For example, 20 fibre-pluggers are needed if the time to move one fibre is 12 seconds. 
A special quality control machine is used to check and clean each module after it has been prepared.
Each module is bar-coded so that the system can keep track of everything.
The modules have to be transported between the telescope and the preparation facility each day.

The choice
between people or robots simply comes down to cost. The cost of employing people is straightforward to calculate.
For example, for 12 seconds per fibre and a total of 7 million spectra we have 23,333 hours of work.
Allowing for rest periods during a shift the total work time for the plugging team could be 30,000 hours.
At a pay rate of
60 euros per hour (equivalent to a cost to ESO of about 100,000 euros per year per person
allowing for weekends and leave) the fibre plugging
team costs a total of 1.8M euros to complete the surveys.
The cost of using robots is much harder to quantify.

The use of plugplates has many adavantages. They have the least possible positional
constraints: there are no magnets or prisms or crossing fibres. There are no positioning
mechanisms attached to the fibres. This is fundamental for scientific versatility! 
Operation during the night is extremely simple and so there is a very low risk of telescope
time being lost compared to systems which reconfigure the fibres on the telescope.
The DE and GA surveys have quite different target density requirements (800 per field for the
high resolution GA survey c.f. up to $\sim$7000 for the high redshift DE survey: both of these
utilise all the detector pixels). Plugplates handle this difference much better than other positioners.
The interface with the spectrographs is simplified (no need for fibre selection mechanisms at the slit end).
All 5 axes for fibre placement (X, Y, Z, tip, tilt) can be controlled if required.
The modules can be very small and light and so they help with the tight mass and space budgets on the VLT.

The reliability of the fibre preparation program is a management issue. By having extra fibre-pluggers, extra
fibre-modules and extra drilling machines, a routine maintenance program will ensure that production
always continues at the required pace.
When fibres break, initially this only impacts on "sky" fibres so the surveys are not affected.
Eventually, once too many fibres are broken in a module it is taken out of operation and repaired.
Similarly, by having spare fibre-pluggers and drilling machines there are always enough operational to meet the survey needs.

\section{Conclusions and discussion}

The DE and GA science cases are extremely attractive. Furthermore, there are other exciting projects
(for example, scaled up versions of projects currently being done with FLAMES) which the facility
envisaged here would be capable of doing.

Optically, a prime focus corrector design with a 1.5 degree FOV is entirely feasible. However, 
the space and mass constraints are tight. An investigation into what drives the mass constraint would be useful - it
may be possible to put more than 1.7 tonnes at prime. Also, the space limitations could be relaxed significantly
by modifying some
parts of the enclosure infrastructure or by using anti-collision interlocks.

The spectrographs need 201 million pixels but are feasible and would probably fit on a single
Nasmyth platform.
The total cost will be substantial (for comparison Gemini have decreed that WFMOS has to cost
less than US\$45M). A proper study is required to establish the cost of a similar facility on the VLT.

If this facility is implemented as a non-interchangable, dedicated facility the
VLT loses 3 instrument stations
and more significantly the option of using VLTI with all 4 UTs for 2-3 years.
On the other hand if the facility is designed to be changed back to the M2
unit within a day then it would not compromise the use of all four UTs for VLTI at all
and only remove one instrument station (assuming
the spectrographs fit on one Nasmyth platform and are never dismounted).



\printindex
\end{document}